\documentclass{aastex62}
\usepackage{amsmath,amssymb,CJK,hyperref}

\received{--}
\revised{--}
\accepted{--}
\submitjournal{ApJ}

\shorttitle{Debris of asteroid disruptions close to the Sun}
\shortauthors{Ye \& Granvik}

\begin{document}
\begin{CJK*}{UTF8}{gbsn}

\title{Debris of asteroid disruptions close to the Sun\footnote{Data and codes that generate the figures and main results of this work are publicly available on\dataset[10.5281/zenodo.2547298]{10.5281/zenodo.2547298} and \url{https://github.com/Yeqzids/near-sun-disruptions}.}}

\correspondingauthor{Quanzhi Ye}
\email{qye@caltech.edu}

\author[0000-0002-4838-7676]{Quanzhi Ye (叶泉志)}
\affiliation{Division of Physics, Mathematics and Astronomy, California Institute of Technology, Pasadena, CA 91125, U.S.A.}
\affiliation{Infrared Processing and Analysis Center, California Institute of Technology, Pasadena, CA 91125, U.S.A.}
\affiliation{Department of Physics and Astronomy, The University of Western Ontario, London, Ontario N6A 3K7, Canada}

\author[0000-0002-5624-1888]{Mikael Granvik}
\affiliation{Division of Space Technology, Lule\aa{} University of Technology, Box 848, 98128 Kiruna, Sweden}
\affiliation{Department of Physics, P.O. Box 64, 00014 University of Helsinki, Finland}



\begin{abstract}
The under-abundance of asteroids on orbits with small perihelion distances suggests that thermally-driven disruption may be an important process in the removal of rocky bodies in the Solar System. Here we report our study of how the debris streams arise from possible thermally-driven disruptions in the near-Sun region. We calculate that a small body with a diameter $\gtrsim0.5$~km can produce a sufficient amount of material to allow the detection of the debris at the Earth as meteor showers, and that bodies at such sizes thermally disrupt every $\sim2$~kyrs. We also find that objects from the inner parts of the asteroid belt are more likely to become Sun-approacher than those from the outer parts. We simulate the formation and evolution of the debris streams produced from a set of synthetic disrupting asteroids drawn from \citet{Granvik2016}'s near-Earth object population model, and find that they evolve 10--70 times faster than streams produced at ordinary solar distances. We compare the simulation results to a catalog of known meteor showers on Sun-approaching orbits. We show that there is a clear overabundance of Sun-approaching meteor showers, which is best explained by a combining effect of comet contamination and an extended disintegration phase that lasts up to a few kyrs. We suggest that a few asteroid-like Sun-approaching objects that brighten significantly at their perihelion passages could, in fact, be disrupting asteroids. An extended period of thermal disruption may also explain the widespread detection of transiting debris in exoplanetary systems.
\end{abstract}



\section{Introduction} \label{sect:intro}

Near-Earth object (NEO) population models have predicted the existence of numerous Sun-approaching asteroids \citep[e.g.][]{Bottke2002, Greenstreet2012}, but after a few decades of rigorous NEO search, few have been found. It has been proposed that thermally-driven ``super-catastrophic'' disruption is responsible for an efficient removal of asteroids that reach a few tenths of an au \citep{Granvik2016}, but such a disruption has not been directly observed.

Rather than catching an asteroid disruption in action, it may be easier to detect the end product of a disruption. A cloud of dust debris resulting from a complete disintegration continues to orbit the Sun on the orbit of the disintegrated asteroid. A range of different effects such as ejection velocity, planetary perturbations and radiation pressure gradually disperse the dust cloud, most noticeably along the orbit, forming a dust (meteoroid) ``stream'' \citep[c.f.][]{Olsson-Steel1987, Brown1998, Williams2004}. The stream continues to disperse over time, eventually blending into the interplanetary meteoroid background \citep[e.g.][]{Cremonese1997e}. Until its dispersal, the meteoroid stream can be detected as a meteor shower on the Earth, if it is on an Earth-crossing orbit and dense enough to stand out from the interplanetary meteoroid background.

Optical and radio meteor surveillance systems, that have been in operation over the past a few decades, have detected a handful of meteoroid streams on Sun-approaching orbits \citep[e.g.][]{Brown2008, Jenniskens2016a}. A few prominent streams, such as the well-known Geminid meteoroid stream, are easily detectable and have been studied for decades \citep[e.g.][and many others]{Denning1893, Plavec1950, Whipple1983, Jones2016, Hui2017, Ryabova2018}. However, most of the Sun-approaching streams are weakly active and have not received a lot of attention \citep{Ye2018}. Many of these streams do not have identifiable parents, raising questions about their formation mechanism.

Here, we present an investigation of the population of Sun-approaching meteoroid streams, with the goal being to critically examine the hypothesis that some (or most) of these streams were produced by thermally-driven disruptions of asteroids with small perihelion distances ($q$). The investigation is divided into two complementary parts: on one hand we will predict the number of small-$q$ meteoroid streams formed by thermally-driven disruptions by utilizing an NEO population model (\S~\ref{sect:model}); on the other hand we will examine the small-$q$ streams that are actually observed (\S~\ref{sect:known}). Results from these two parts will be compared to each other and discussed in \S~\ref{sect:disc}.

\section{Predicted Characteristics of Thermally-Driven Streams} \label{sect:model}

\subsection{Size of the Parent and the Detectability of the Resulting Meteoroid Streams\footnote{The Jupyter notebook that shows the calculation of the numbers mentioned in this section can be found \href{https://github.com/Yeqzids/near-sun-disruptions/blob/master/nb/stream_mass.ipynb}{here}.}} \label{sect:model:size}

Meteor showers are essentially local overdensities of meteor radiants; therefore, meteoroid streams that are detectable need to stand out against the interplanetary meteoroid background. Since the flux intensity of a stream is directly related to the production rate of the parent, a disrupting parent needs to be massive enough to produce enough dust to supply a detectable stream. Therefore, the first question is the critical parent size needed to produce a detectable stream.

Radar and optical techniques are the most widely used methods to detect meteors \citep{Jenniskens2017}, therefore, in this work, we focus on the regimes explored by these two techniques. The detection limit of typical meteor radars are in the range of $\sim10^{-3}$ to $10^{-2}~\mathrm{km^{-2}~hr^{-1}}$, appropriate to meteoroid sizes down to 1~mm \citep{Ye2016,Ye2016b}, while for typical video systems, $10^{-5}$ to $10^{-4}~\mathrm{km^{-2}~hr^{-1}}$, appropriate to meteoroid sizes down to 1~cm \citep{Jenniskens2016a,Jenniskens2016}. These two numbers translate to a Zenith Hourly Rate (ZHR) of $\lesssim1$ using the relation derived by \citet[][\S~10]{Koschack1990}, assuming a typical duration of stream activity of a couple days, collection area (i.e., the area of atmosphere that one system monitors) of $10^4~\mathrm{km^2}$, a typical meteoroid mass index of $\sim1.6$ \citep{Blaauw2011}, and system uptime to be 6~hr each day, which is in the same ballpark as the typical background flux observed by visual observers.

The mass of a meteoroid stream can be calculated following the derivation of \citet{Hughes1989}:

\begin{equation}
\label{eq:msmass}
M = \frac{f \pi t^2 V_\mathrm{E}^2 I \sin{^2\epsilon} V_\mathrm{H} P}{4 V_\mathrm{G}}
\end{equation}

\noindent where $f\sim10$ is a dimensionless factor that accounts for the shape of the stream \citep{McIntosh1988}, $t$ is the duration of the meteor shower at the Earth in seconds, $V_\mathrm{E}=29700~\mathrm{m~s^{-1}}$ is the orbital speed of the Earth, $I$ is the mass influx at the Earth in $\mathrm{kg~m^{-2}~s^{-1}}$, $\epsilon$ is the angle between the Earth's path and the orbit of the stream, $V_\mathrm{H}$, $V_\mathrm{G}$ is the heliocentric and geocentric speed of the meteoroids, respectively, and $P$ is the orbital period of the meteoroids in seconds. Recognizing that such a calculation is only to be taken at an order of magnitude level, we take $\sin{\epsilon} \sim 1$, $V_\mathrm{H}/V_\mathrm{G} \sim 1$, average meteoroid density of $2000~\mathrm{kg~m^{-3}}$ \citep{Rotundi2015}, and the flux and meteoroid sizes discussed above, we derive $M\sim10^{11}$~kg as the minimum mass of the progenitor that is needed for the detection of the resulting meteoroid stream by contemporary meteor-detection networks. This number corresponds to a progenitor diameter of $\sim0.5$~km \citep[assuming a density of $2000~\mathrm{kg~m^{-3}}$, c.f.][]{Carry2012}. The stream mass derived from different observations can differ by a factor of 10 from the mean \citep{Ryabova2017}, which translates to a factor of $\sim2$ in the uncertainty of the progenitor size.

\subsection{Rate of Thermally-Driven Disruptions} \label{sect:model-rate}

How often does thermally-driven disruption occur? With some necessary simplification of the problem, this can be estimated using an NEO population model. Here we use the \citet{Granvik2018} model which considered the effect of thermally-driven disruption of NEOs. We start from the predicted flux of NEOs from known sources of NEOs in the main asteroid belt, and then multiply these fluxes with the predicted likelihood that an asteroid from a specific region of the main asteroid belt would reach the critical heliocentric distance from the Sun \citep[0.058~au for km-sized asteroids, as derived by][]{Granvik2016}. Finally, we sum up the products to arrive at a total rate of thermal disruptions of $550\pm30\,\mathrm{Myr}^{-1}$, where the error accounts for uncertainties in the flux of NEOs from different parts of the asteroid belt.

The result of our calculation is summarized in Table~\ref{tbl:model-rate}. We observe two things: (1) asteroids from the inner main-belt region (Hungarias, Phocaeas, low-inclination $\nu_6$ objects) are more likely to become Sun-approachers; and (2) asteroids from the inner main-belt region also tend to spend a longer time as Sun-approachers compared to outer main-belt asteroids. A direct implication of these two features is that meteoroid streams from the inner belt should dominate the thermally-driven streams, while streams with semimajor axes $a$ compatible with an origin in the outer belt are statistically unlikely to be thermally-driven disrupted asteroids.

\begin{table*}
\begin{center}
\caption{Probabilities that asteroids from different escape regions in the asteroid belt end up on Sun-approaching orbits ($P_\mathrm{sungrazer}$) as well as timescale that they have $q<0.14$~au prior to disruption at $q=0.058$~au ($t_\mathrm{sungrazer}$). Escape regions with the format of $X$:$Y$J stand for $X$:$Y$ mean-motion resonance with Jupiter. \label{tbl:model-rate}}
\begin{tabular}{ccc}
\hline
Escape region & $P_\mathrm{sungrazer}$ & $t_\mathrm{sungrazer}$ \\
\hline
Hungaria & 78\% & 13000~yr \\
Phocaea & 89\% & 1400~yr \\
$\nu_6$ complex & 81\% & 13000~yr \\
3:1J complex & 74\% & 2800~yr \\
5:2J complex & 22\% & 1500~yr \\
2:1J complex & 22\% & 700~yr \\
\hline
\end{tabular}
\end{center}
\end{table*}

\subsection{Distribution and Behavior of Thermally-Produced Streams} \label{sect:model:sim}

\subsubsection{Simulation Setup}

\begin{figure*}
\includegraphics[width=\textwidth]{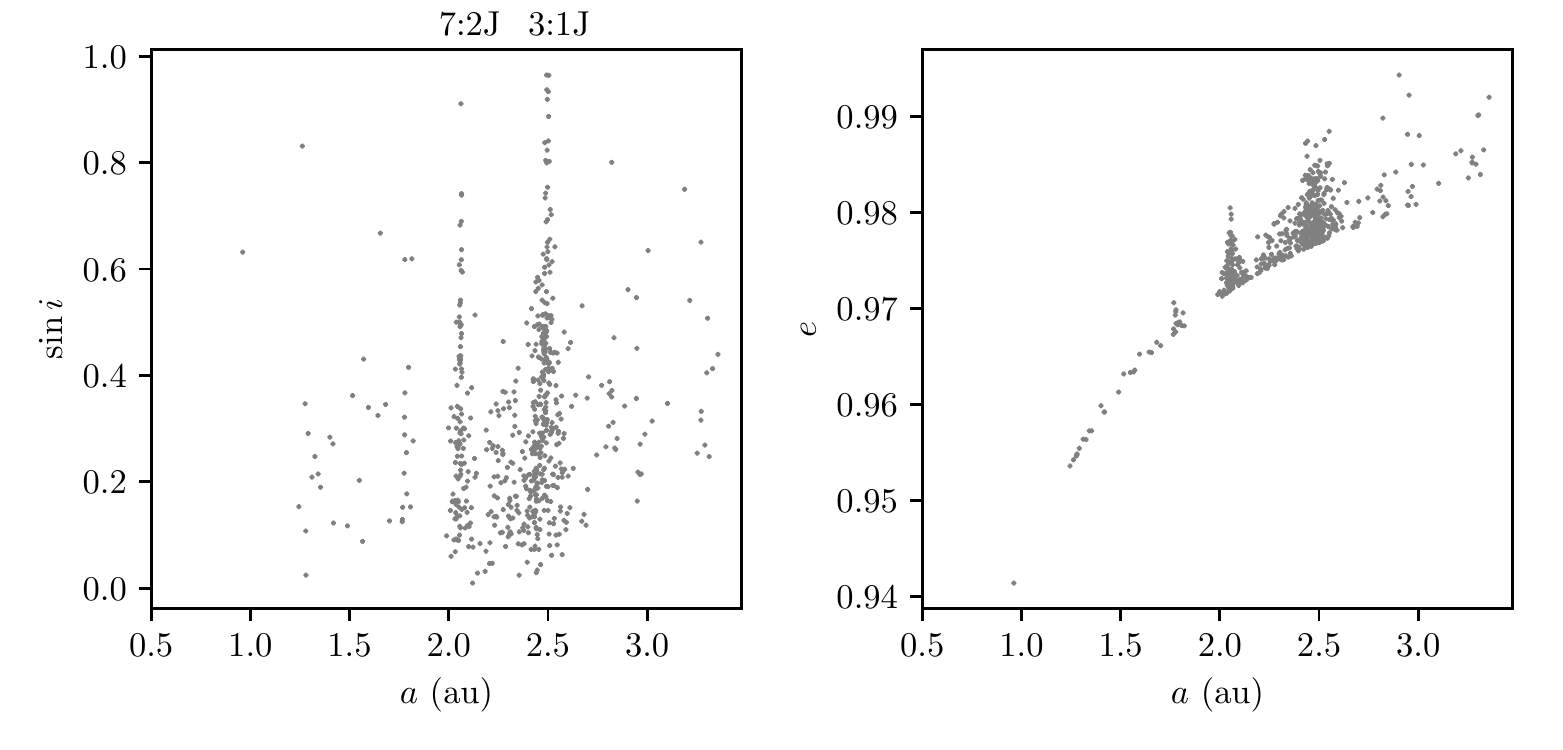}
\caption{$a$, $e$ and $\sin{i}$ of the synthetic asteroids drawn from \citet{Granvik2016}'s NEO population model. The overdensities produced by 7:2J (which coincides with Hungaria, Phocaea, and $\nu_6$ at low inclinations) and 3:1J resonance are marked. The Jupyter notebook for this figure is available \href{https://github.com/Yeqzids/near-sun-disruptions/blob/master/nb/test_asteroids.ipynb}{here}. \label{fig:neo-model}}
\end{figure*}

We randomly select 1\% (608 out of 60,727) of the test asteroids in the orbital integrations by \citet{Granvik2016} that have perihelion distances within the critical disruption distance of $q_\ast=0.058$~au, representing the population of thermally disrupting asteroids (Figure~\ref{fig:neo-model}). We note that the ($a,\sin{i}$) distribution covers the essential parts of the phase space whereas the ($a,e$) distribution is limited to orbits with small perihelion distances, as it should be. We then generate a total of 500 particles for each test asteroid at their respective perihelion point, mimicking the thermally-driven disintegration of asteroids. The size distribution of these particles follows a size range of $a \subseteq [5\times10^{-4}, 5\times10^{-2}]$~m, a meteoroid bulk density of $2000~\mathrm{km~m^{-3}}$, and a continuous size distribution of $\mathrm{d}N/\mathrm{d}a \propto a^{-q}$ where $q=3.6$ as suggested by telescopic observations \citep[e.g.][]{Fulle2004a, Ye2016a}.

\begin{figure*}
\includegraphics[width=\textwidth]{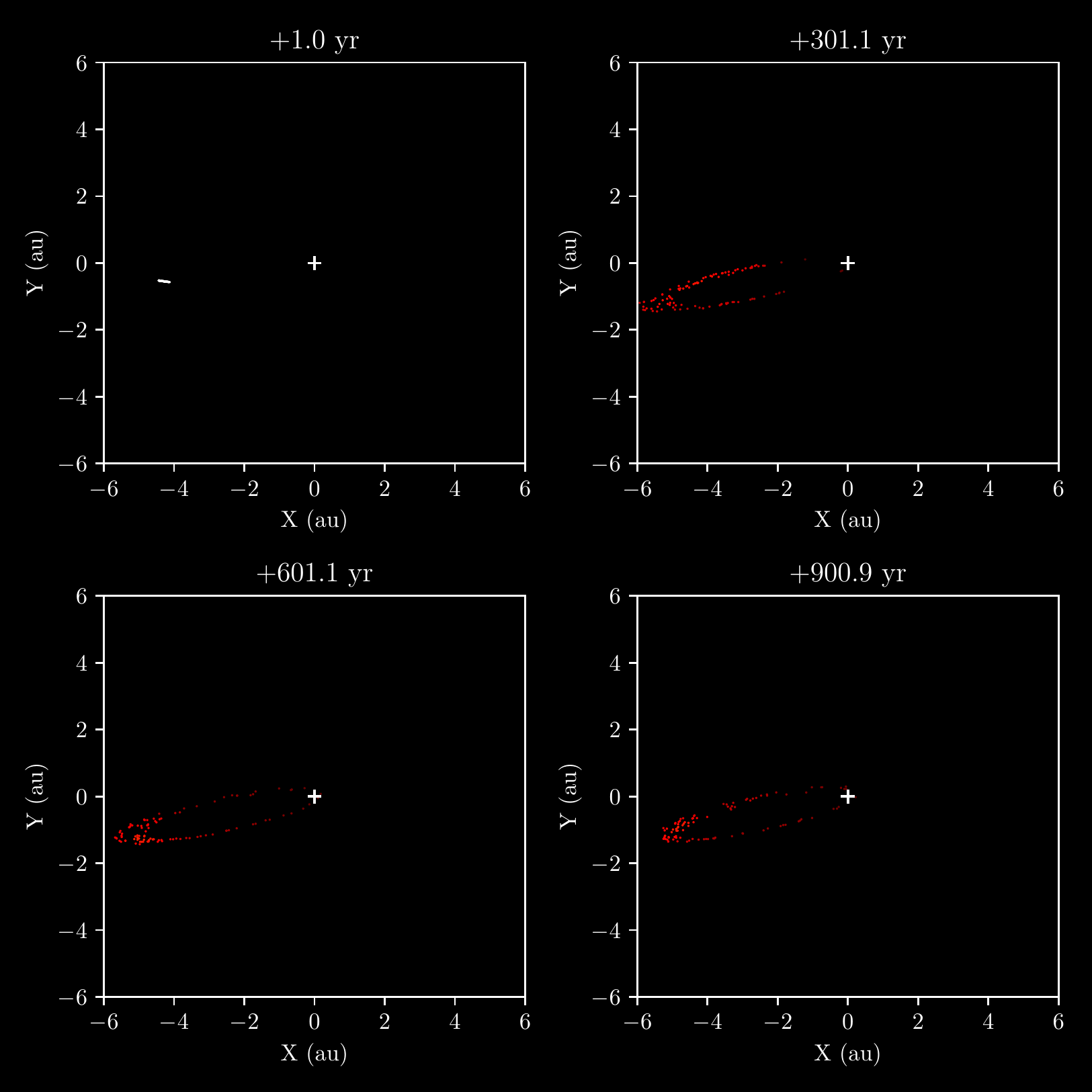}
\caption{Disruption of four randomly selected synthetic asteroids, assuming dust ejection at gravitational escape speed of a km-sized body. The Sun is marked by a white cross. The Jupyter notebook for this figure/animation is available \href{https://github.com/Yeqzids/near-sun-disruptions/blob/master/nb/test_asteroids.ipynb}{here}. \label{fig:stream-sim}}
\end{figure*}

We integrate these sub-particles for 10~kyr using the Bulirsch-Stoer integrator \citep{Stoer2013} embedded in a tailored MERCURY6 $N$-body simulator \citep{Chambers1999,Ye2016}. The code accounts for gravitational perturbations by major planets (Mercury through Neptune with the Earth-Moon system represented by a single point of mass), radiation pressure, Poynting-Robertson drag, as well as perihelion shift due to general relativity. The orbital elements of all test particles are recorded with a cadence of 1~yr. The animation in Figure~\ref{fig:stream-sim} depicts the formation, evolution and final dispersion of a meteoroid stream.

The simulation is run two times with different ejection model: the first one with \citet{Crifo1997}'s cometary ejection model\footnote{Readers may wonder about choosing \citet{Crifo1997}'s model over the \citet{Whipple1951}'s model. These two models are derived from different approaches but produced similar end results. As we will show in the following section, the choice of ejection model does not have significant impact on our conclusion, and is hence unimportant.}; and the second one with an ejection model that assumes gravitational escape ejection (i.e. eject at escape velocity). This is due to the poorly understood ejection scheme of a thermally-driven disruption, but it is reasonable to expect that the ejection speed should be somewhere between the one set by cometary model and the one set by gravitational escape model. Here we note that C/2015 D1 (SOHO), a Sun-approaching comet that disintegrated during its perihelion passage in early 2015 with excess thermal stress as the likely cause, showed morphology consistent with comet-like ejection \citep{Hui2015}; however, C/2015 D1 is of cometary origin and the lesson it provided may not be applicable to asteroids.

We also acknowledge that the information provided by 500 particles per test asteroid is somewhat limited, considering that many meteoroid stream simulations account for thousands of particles. We are mainly limited by the high computational cost of the simulation: the full simulation of 608 meteoroid streams over 10~kyrs with two sets of input parameters takes about 50 CPU years on a single 2.2~GHz AMD Opteron CPU, though with distributed computing we are able to complete our simulation within a few months. The exact numbers derived for each stream become less important as we collectively examine a large set of streams, which we believe is sufficient for deriving a broad, global picture of asteroid disruptions close to the Sun.

\subsubsection{Calculation of Stream Formation, Dispersion, and Visible Timescale}

Following the definition in our earlier work \citep{Ye2016}, we define the stream formation time as the time taken to the point that the standard deviation of the mean anomalies of the test particles reach $60^\circ$ (i.e. the $3\sigma$ limits cover the entire orbit if the distribution of mean anomalies is Gaussian), while the stream-dispersal timescale is defined as the time it takes for $50\%$ of the test particles to lose stream coherency. The stream coherency is defined using the \citet{Southworth1963}'s decoherence parameter ($D$), which is defined as:

\begin{equation}
 D_{A, B}^2 = \left(q_B - q_A \right)^2 + \left( e_B - e_A \right)^2 + \left( 2\sin{ \frac{I}{2} } \right)^2 + \left[ \left(e_A + e_B \right) \sin{ \frac{\varPi}{2} } \right]^2
\end{equation}
where
\begin{equation}
 I = \arccos{ \left[ \cos{i_A} \cos{i_B} + \sin{i_A} \sin{i_B} \cos{\left( \varOmega_A - \varOmega_B \right)} \right]} 
\end{equation}
and
\begin{equation}
 \varPi = \omega_A - \omega_B + 2 \arcsin{ \left( \cos{\frac{i_A+i_B}{2}} \sin{\frac{\varOmega_A-\varOmega_B}{2}} \sec{\frac{I}{2}} \right) } 
\end{equation}

\noindent and the subscripts $A$ and $B$ refer to the two test particles being compared, $q$ is the perihelion distance in au, $e$ is the eccentricity, $i$ is the inclination, $\Omega$ is the longitude of ascending node, and $\omega$ is the argument of perihelion. The sign of the $\arcsin$ term in the equation for $\varPi$ switches over if $|\Omega_A-\Omega_B|>180^\circ$. We adopt an empirical cutoff of $D=0.1$ used by many of the past works \citep[e.g.][]{Drummond1981, Jopek1993}, noting that cutoffs found by more rigorous tests \citep[e.g.][]{Fu2005a, Moorhead2016} are not substantially different from $D=0.1$.

Streams cannot be detected unless they intercept the Earth's orbit, therefore we also need to calculate their ``visible time'' at Earth. A stream is considered to be ``visible'' if the spread of either or both of the heliocentric distances of their ascending/descending nodes encompass 1~au (Earth's average distance to the Sun) before it disperses. We then sum up the time that stream is visible to derive the ``visible timescale'' at the Earth.

\subsubsection{Results} \label{sect:model:result}

We find that results from different ejection models are not dramatically different: the ``cometary'' model predicts a median formation and dispersion timescale of $6\pm1$~yr and $170_{-130}^{+1500}$~yr, while the gravitational escape model predicts $8\pm1$~yr and $280_{-240}^{+2000}$~yr; both ejection models predict that $\sim75\%$ of the simulated streams will be visible at some point during their lifetime, for a median total time of $40_{-25}^{+140}$~yr\footnote{The Jupyter notebook for this calculation is available \href{https://github.com/Yeqzids/near-sun-disruptions/blob/master/nb/formation_and_dispersion_timescales.ipynb}{here}.}. (Error bars indicate $1\sigma$ probability interval.) The formation and dispersion timescales are significantly shorter than for typical (non-Sun-approaching) meteoroid streams, which take $400\pm80$~yr to form and $3000\pm300$~yr to disperse \citep{Ye2016}.

\begin{figure*}
\includegraphics[width=\textwidth]{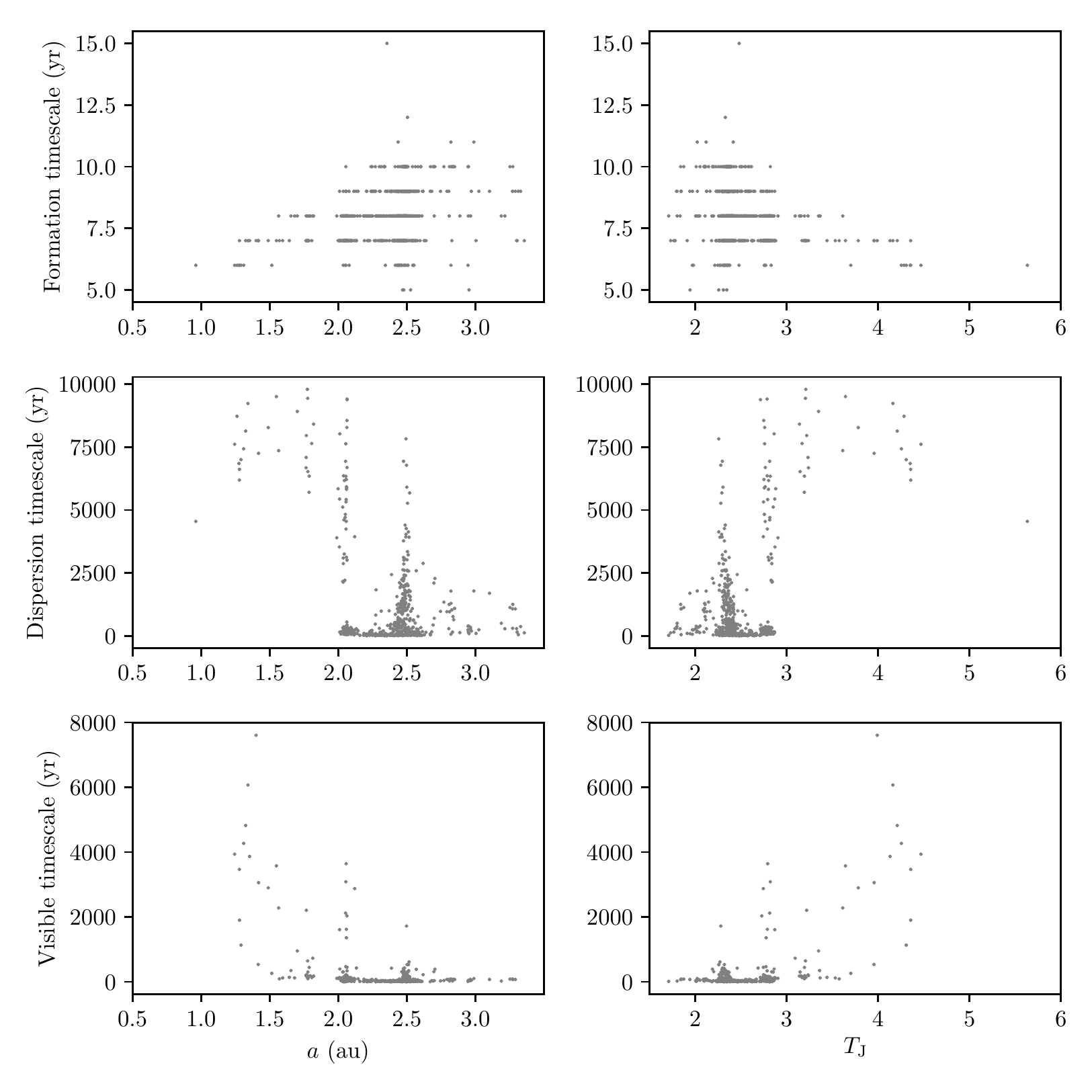}
\caption{Different timescales of the simulated meteoroid streams as well as their dependence on $a$ and $T_\mathrm{J}$ of the streams. Distributions are drawn from the simulations that assume ejection at gravitational escape speed. The formation timescales (the two upper panels) look discrete as they are comparable to the output interval of the integration (1~yr). The Jupyter notebook for this figure is available \href{https://github.com/Yeqzids/near-sun-disruptions/blob/master/nb/timescales_a_tj.ipynb}{here}. \label{fig:time-a-tj}}
\end{figure*}

Figure~\ref{fig:time-a-tj} shows the dependence of various timescales to orbital elements $a$ and $T_\mathrm{J}$. No clear dependence is seen between the formation timescale and dynamical properties of the parent. The dispersion and visible timescales, on the other hand, show a clear dependence on whether the stream is on a Jupiter-approaching orbit (which can be measured by $T_\mathrm{J}$) or close to resonances. Streams that are dynamically decoupled from Jupiter (i.e. $T_\mathrm{J}>3$) are longer-lived than those that are not, except for the streams that originate from parents close to resonances (Hungaria, Phocaea, 7:2J, or 3:1J complex).

One key takeaway from Figure~\ref{fig:time-a-tj} is that thermally-produced meteoroid streams have very short visibility at the Earth, due to the fact that they are mostly shortly-lived and are dynamically quickly evolving. Given a typical visible timescale of $40$~yr, if we consider the disruption rate of km-sized NEOs that we derived in \S~\ref{sect:model-rate}, which is $\sim0.6$~kyr$^{-1}$, we can immediately see that the chance of seeing a thermally-produced meteoroid stream is close to zero.

\section{Known Sun-approaching Meteoroid Streams} \label{sect:known}

The official catalog of meteor showers is managed by the International Astronomical Union's Meteor Data Center \citep[IAUMDC;][]{Jopek2011,Jopek2014,Jopek2017}. Under the current rule of meteor shower nomenclature, a previously unreported stream will be added to the Working List of Meteor Showers, but will only be considered for ``established'' status when it is independently re-detected by other observers \citep{Jopek2014, Janches2018}. Promotion of status and assignment of official names is overseen by the IAU F1 Commission during the triennial General Assemblies (GA) meeting. Currently, the most recent version of the IAUMDC catalog was reviewed at GA meeting in August 2018 and dated 2018 November 9. This version has a total of 932 meteor showers with 112 of them considered to be established and will be used in the analysis in this work. Since showers on the Working List are to be considered provisional detections and thus subjected to removal if they cannot be confirmed, we will focus on the established showers for now.

Table~\ref{tbl:shr} lists all 22 established Sun-approaching meteoroid streams in the IAUMDC catalog that have perihelion distance $q<0.15$~au. A loose cutoff at 0.15~au is chosen because the Geminids/GEM\footnote{For readability, from now on we will use the IAUMDC three-letter code to refer to a shower. The full name of the shower is listed in Table~\ref{tbl:shr}.} stream, the prominent candidate for thermally-produced streams, has $q=0.14$~au. The 22 streams have the following characteristics:

\begin{enumerate}
 \item 8 have $T_\mathrm{J}<2$ (i.e. compatible with orbits of Halley-type and long-period comets). Even though only one of them has a proposed parent body which is a comet \citep[KLE and C/1917 F1, c.f.][]{Veres2011,Neslusan2014}, they are most likely to be produced by comets based on their dynamical properties.
 \item 4 have $2<T_\mathrm{J}<3$. This type of orbit is compatible with Jupiter-family comets and asteroidal interlopers \citep{Levison1994,Jewitt2015,Hsieh2016}. Among these showers, SDA has a proposed parent (the 96P/Machholz complex) that is considered to be rather definitive \citep[e.g.][]{Abedin2017,Abedin2018}.
 \item 10 have $T_\mathrm{J}>3$ that are dynamically in the asteroid regime. The 96P/Machholz Complex is thought to be responsible for ARI \citep{Bruzzone2014,Abedin2017,Abedin2018}, a stream that is near the borderline of $T_\mathrm{J}=3$, while (3200) Phaethon and (155140) 2005 UD are widely believed to be the parents of the Geminids and Daytime Sextantids. IAUMDC lists 3 comets found by the Solar and Heliospheric Observatory (SOHO) probe as likely parents for NOC, NZC and OCE, but no reference is provided and we cannot find any published papers that discussed these linkages; additionally, the orbits of these SOHO comets are poorly known due to the extremely short arc (less than a day for two of the comets, a week for the third), therefore the linkages are doubtful. In either case, there are 4--7 dynamically asteroidal streams that have no known parent bodies associated with them.
\end{enumerate}

\begin{table*}
\tiny
\begin{center}
\caption{Established showers in the IAUMDC database with $q<0.15$~au. Listed parameters are IAUMDC's three-letter code and name, solar longitude of peak activity, radiant in J2000 reference frame and geocentric speed at peak activity, orbital elements including $a$, $e$, $i$, and $T_\mathrm{J}$, the technique that detected the shower (O = optical, R = radar), the estimated age from \S~\ref{sect:known:age-estimate} (if available), and proposed parent body in the IAUMDC catalog. The data spreadsheet is available \href{https://github.com/Yeqzids/near-sun-disruptions/blob/master/data/streamfulldata_established.csv}{here} and a Jupyter notebook that used to prepare this table available \href{https://github.com/Yeqzids/near-sun-disruptions/blob/master/nb/shower_table_established.ipynb}{here}. \label{tbl:shr}}
\begin{tabular}{lccccccccccl}
\hline
IAUMDC code/name & $\lambda_\odot$ & \multicolumn{2}{c}{Radiant} & $V_\mathrm{G}$ & $a$ & $e$ & $i$ & $T_\mathrm{J}$ & Tech. & Est. age & Parent body \\
 & & RA & Decl. & (km/s) & (au) & & & & & (kyr) & \\
\hline
AAN $\alpha$ Antliids & 313.1$^\circ$ & 160.7$^\circ$ & -11.9$^\circ$ & 43.9 & 2.4 & 0.94 & 62.7$^\circ$ & 2.5 & O+R & 4 & \\
ARI Daytime Arietids & 76.2$^\circ$ & 42.5$^\circ$ & +24.0$^\circ$ & 38.2 & 1.9 & 0.95 & 24.6$^\circ$ & 3.2 & O+R & 5 & 96P/Machholz complex \\
CTA $\chi$ Taurids & 220.5$^\circ$ & 63.1$^\circ$ & +25.4$^\circ$ & 41.6 & 4.9 & 0.98 & 13.7$^\circ$ & 1.4 & O+R & - & \\
DLT Daytime $\lambda$ Taurids & 85.5$^\circ$ & 56.7$^\circ$ & +11.5$^\circ$ & 36.4 & 1.6 & 0.93 & 23.2$^\circ$ & 3.7 & O+R & 5.5 & \\
DSX Daytime Sextantids & 186.7$^\circ$ & 155.0$^\circ$ & -1.6$^\circ$ & 31.8 & 1.1 & 0.86 & 22.5$^\circ$ & 5.1 & O+R & 6-7 & (155140) 2005 UD \\
EPG $\epsilon$ Pegasids & 108.6$^\circ$ & 329.9$^\circ$ & +14.5$^\circ$ & 28.6 & 0.7 & 0.78 & 49.7$^\circ$ & 7.3 & O+R & 6-9 & \\
EPR $\epsilon$ Perseids & 91.1$^\circ$ & 55.7$^\circ$ & +37.6$^\circ$ & 44.3 & 7.3 & 0.98 & 57.1$^\circ$ & 1.0 & O+R & - & \\
GEM Geminids & 261.6$^\circ$ & 113.0$^\circ$ & +32.3$^\circ$ & 34.5 & 1.4 & 0.90 & 23.5$^\circ$ & 4.2 & O+R & 6.5-7 & (3200) Phaethon \\
JLE January Leonids & 282.5$^\circ$ & 148.1$^\circ$ & +23.9$^\circ$ & 52.1 & 5.7 & 0.99 & 105.8$^\circ$ & 0.8 & O+R & - & \\
KLE Daytime $\kappa$ Leonids & 182.1$^\circ$ & 162.2$^\circ$ & +15.3$^\circ$ & 43.4 & 20.2 & 0.99 & 25.0$^\circ$ & 0.9 & R & - & C/1917 F1 (Mellish) \\
NDA Northern $\delta$ Aquariids & 140.6$^\circ$ & 345.6$^\circ$ & +1.0$^\circ$ & 39.2 & 2.2 & 0.96 & 22.0$^\circ$ & 2.8 & O+R & 4-5 & \\
NOC Northern Daytime $\omega$ Cetids & 46.6$^\circ$ & 5.7$^\circ$ & +17.6$^\circ$ & 34.9 & 1.3 & 0.91 & 38.1$^\circ$ & 4.7 & O+R\tablenotemark{a} & 6.5-7 & See note\tablenotemark{b} \\
NOO November Orionids & 245.9$^\circ$ & 89.1$^\circ$ & +15.4$^\circ$ & 43.1 & 11.2 & 0.99 & 36.2$^\circ$ & 0.8 & O+R & - & \\
NZC Northern June Aquilids & 99.1$^\circ$ & 308.4$^\circ$ & -5.1$^\circ$ & 37.8 & 1.7 & 0.93 & 38.8$^\circ$ & 3.4 & O+R & 6 & See note\tablenotemark{c} \\
OCE Southern Daytime $\omega$ Cetids & 46.4$^\circ$ & 20.7$^\circ$ & -5.6$^\circ$ & 36.7 & 1.6 & 0.92 & 35.1$^\circ$ & 3.5 & R & 6 & See note\tablenotemark{b} \\
PAU Piscis Austrinids & 131.2$^\circ$ & 350.2$^\circ$ & -22.1$^\circ$ & 44.0 & 4.4 & 0.97 & 58.6$^\circ$ & 1.5 & O+R & - & \\
SDA Southern $\delta$ Aquariids & 126.8$^\circ$ & 336.4$^\circ$ & -16.1$^\circ$ & 40.9 & 2.6 & 0.97 & 28.7$^\circ$ & 2.3 & O+R & 4 & 96P/Machholz complex \\
SSE $\sigma$ Serpentids & 275.5$^\circ$ & 243.5$^\circ$ & -1.7$^\circ$ & 43.5 & 2.7 & 0.93 & 62.1$^\circ$ & 2.4 & O+R & 4 & \\
SZC Southern June Aquilids & 88.2$^\circ$ & 307.3$^\circ$ & -31.4$^\circ$ & 37.0 & 1.4 & 0.93 & 41.9$^\circ$ & 4.2 & O+R & 3.5-7 & \\
THA November $\theta$ Aurigids & 240.5$^\circ$ & 92.3$^\circ$ & +34.7$^\circ$ & 33.1 & 1.1 & 0.89 & 26.4$^\circ$ & 5.0 & O+R & 4-8 & \\
XRI Daytime $\xi$ Orionids & 123.7$^\circ$ & 98.7$^\circ$ & +15.9$^\circ$ & 42.6 & 6.8 & 0.98 & 27.3$^\circ$ & 1.1 & R & - & \\
ZCA Daytime $\zeta$ Cancrids & 153.5$^\circ$ & 127.9$^\circ$ & +15.3$^\circ$ & 43.0 & 4.8 & 0.99 & 18.9$^\circ$ & 1.4 & R & - & \\
\hline
\end{tabular}
\end{center}
\tablenotetext{a}{The IAUMDC catalog (version 2018 November 9) listed only radar detection of this shower, but optical detection has been reported by \citet{Jenniskens2018}.}
\tablenotetext{b}{The IAUMDC catalog gives C/2003 Q1 (SOHO) as a likely parent, but we find no published research that proposed or discussed this linkage. The orbit of C/2003 Q1 (SOHO) is based on extremely short arc (less than a day), therefore the linkage is questionable.}
\tablenotetext{c}{The IAUMDC catalog gives C/1997 H2 (SOHO) and C/2009 U10 (SOHO) as likely parents, but we find no published research that proposed or discussed this linkage. Orbits of these two comets are based on very short arcs (6 days for C/1997 H2, 1 day for C/2009 U10). The linkages are therefore questionable.}
\end{table*}

\subsection{Ages of Meteoroid Streams}

\subsubsection{Age-Width-$T_\mathrm{J}$ Map} \label{sect:age:map}

Aside from the orbit, which gives us an idea of the dynamical properties of a stream, another piece of useful information that we can get from meteor observations is the age of the stream. As noted in \S~\ref{sect:intro}, the stream age correlates to the dispersion of the stream, which can be measured through the width of the stream as the Earth travels through the orbital intersection, i.e., the duration of the shower activity. Although an accurate determination of the age requires knowledge of a number of poorly constrained parameters, such as how meteoroids are ejected as well as their physical and mechanical properties, and is therefore difficult if not impossible to carry out, a crude comparison between the observed shower width and dynamical models is usually sufficient to broadly constrain the age of a stream. This has been done for many meteoroid streams including Quadrantids, Perseids, Geminids and others \citep[e.g.][]{Williams1993,Brown1998, Ryabova1999,Abedin2018}.

However, this method requires \textit{a priori} knowledge of the orbit of the parent, since the dynamics of the stream and the parent evolve over time and are not fully correlated with each other. Most of the streams in Table~\ref{tbl:shr} do not have known parents and, as we have shown earlier, Sun-approaching streams evolve faster than typical meteoroid streams. Therefore we must seek a different path to achieve our goal.

Our solution is to reuse the $N$-body simulations completed in \S~\ref{sect:model} to map the dependence between age, stream width, and $T_\mathrm{J}$. $T_\mathrm{J}$ is one of the variables here since the perturbation from Jupiter is a dominant factor of the dynamical evolution of meteoroid streams. The major benefit of this method is that it bypasses the need to know the parent's orbit, since the set of possible orbits of the parent is already captured by the NEO population model.

We focus on the 14 $T_\mathrm{J}>2$ streams in Table~\ref{tbl:shr}, as we are most interested in the thermally-driven disruptions of asteroids and the NEO population model is not applicable to $T_\mathrm{J}<2$ objects. To generate the age-width-$T_\mathrm{J}$ map, we first determine the width of the stream and the mean $T_\mathrm{J}$ for each simulated stream at each time step. The stream width is defined as the $90\%$ width of the longitude of the ascending node ($\Omega$), the point where the particle passes the ecliptic plane (i.e. the plane of Earth's orbit). The $90\%$ percentile is used to reject the few random particles that might have been gravitationally scattered during the integration. Mean $T_\mathrm{J}$ is defined as the mean of the $T_\mathrm{J}$ values of all particles. We calculate the mean stream width and $T_\mathrm{J}$ for every time step until the stream has lost $50\%$ of the particles (due to solar impact or ejection from the Solar System) or the end of the integration has been reached. To increase the clarity of the map, we apply a Gaussian filter with $\sigma=3.0$ (an arbitrarily chosen number) to the derived age-width-$T_\mathrm{J}$ map to remove sharp gradients resulting from limited statistics.

\begin{figure*}
\includegraphics[width=\textwidth]{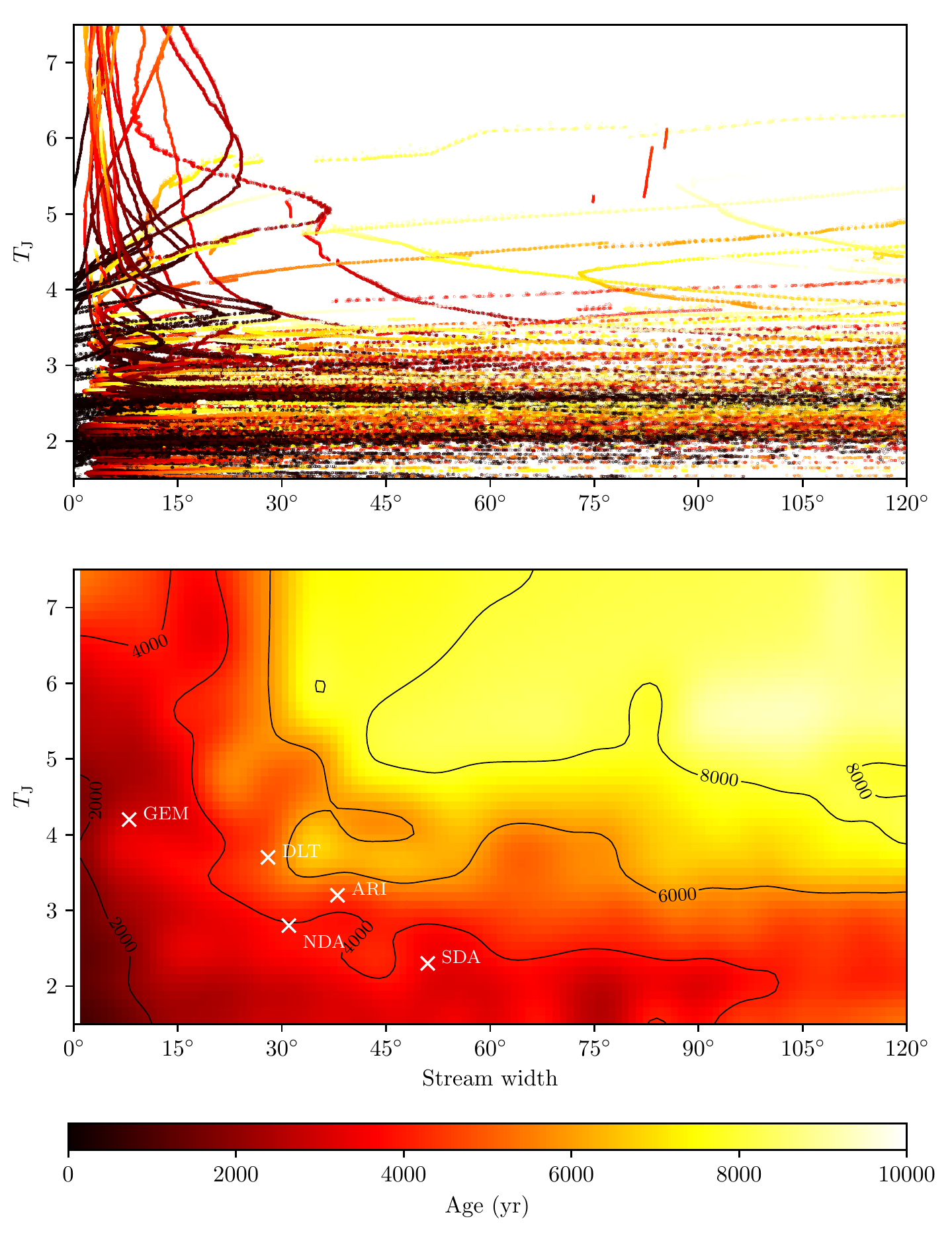}
\caption{Raw (top) and smoothed (bottom) age-width-$T_\mathrm{J}$ map using cometary ejection model. The Jupyter notebook for this figure is available \href{https://github.com/Yeqzids/near-sun-disruptions/blob/master/nb/age_map.ipynb}{here}. \label{fig:age-width-tj-cmt}}
\end{figure*}

\begin{figure*}
\includegraphics[width=\textwidth]{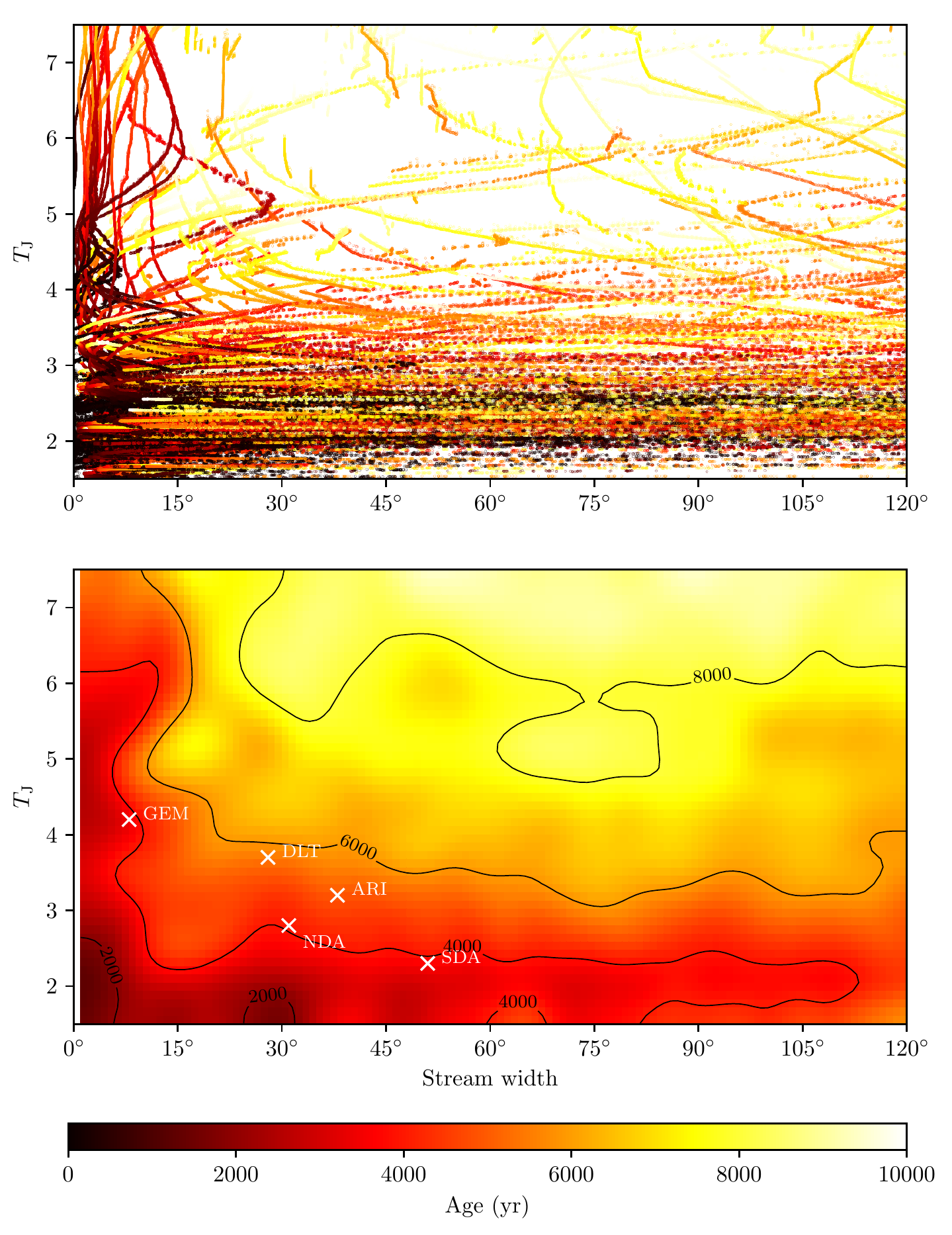}
\caption{Raw (top) and smoothed (bottom) age-width-$T_\mathrm{J}$ map using gravitational escape ejection model. The Jupyter notebook for this figure is available \href{https://github.com/Yeqzids/near-sun-disruptions/blob/master/nb/age_map.ipynb}{here}. \label{fig:age-width-tj-grav}}
\end{figure*}

The aforementioned procedure is applied to the simulation results obtained from both the cometary ejection model and the gravitational escape ejection model. The raw input as well as the final processed age-width-$T_\mathrm{J}$ map are shown in Figures~\ref{fig:age-width-tj-cmt} and \ref{fig:age-width-tj-grav}. Our primary goal here is to validate the accuracy of this map and use it to estimate the ages of the streams in Table~\ref{tbl:shr}, but prior to validation we do note the following general features of the map:

\begin{enumerate}
 \item As expected, streams at higher $T_\mathrm{J}$ are generally longer-lived than those at lower $T_\mathrm{J}$.
 \item Broadly speaking, there is a visible albeit not dramatic difference between the maps derived from cometary ejection model and gravitational escape ejection model, consistent with the numbers we derived in \S~\ref{sect:model}.
 \item The raw inputs (the upper panel of each figure) show broken short segments: this is because some evolved streams have $>10\%$ particles scattered into very different orbits and therefore bloat the stream width. The stream width shrinks once these highly unstable particles are ejected from the Solar System.
 \item The raw inputs also show appreciable scatters of age values across the $T_\mathrm{J}$-width space, which seems to undermine our method. However, we note that neither $T_\mathrm{J}$ nor stream width can be tightly constrained from observations, and that our goal is to identify broad ranges of evolutionary ages compatible with the observations. The validation, to be described below, confirms that our estimates are broadly consistent with the ones derived from stream-specific models. Therefore, the scattering is not of a concern for our purpose.
 \item A small but non-negligible fraction of the streams have evolved from low $T_\mathrm{J}$ to high $T_\mathrm{J}$, which we will discuss later.
\end{enumerate}

Dynamics of several streams in Table~\ref{tbl:shr} have been previously studied by other researchers, providing an opportunity for a sanity check. These streams are marked in both figures: \citet{Abedin2017} and \citet{Abedin2018} reported the age of ARI, DLT, NDA and SDA (all originated from the 96P/Machholz complex) to be 10--20~kyr; various studies on GEM suggested an age of a few kyr \citep[c.f.][]{Nesluvsan2015}.

We mark the locations of ARI, DLT, NDA, SDA and GEM on Figures~\ref{fig:age-width-tj-cmt} and \ref{fig:age-width-tj-grav}, using the stream width used by the original research \citep[for GEM, we use the most recent visual data that has been used for some of the recent modeling work, c.f.][]{Arlt2006, Ryabova2016} and $T_\mathrm{J}$ calculated from the mean orbit provided by IAUMDC.

At the first glance, estimates using our map agree well on the age of the Geminids but underestimate the age of the 96P/Machholz streams. The main reason for that is that \citet{Abedin2017} and \citet{Abedin2018} assumed continuous ejection from 96P/Machholz, while we assume one-time ejection to mimic thermal disruption events. Assuming constant ejection rate and meteoroid delivery efficiency, the age derived from continuous ejection model should be twice as long as one-time ejection model, which suggests that our estimate is, in fact, in line with the numbers derived by \citet{Abedin2017} and \citet{Abedin2018}.

\subsubsection{Estimating the Age} \label{sect:known:age-estimate}

To estimate the age using the age-width-$T_\mathrm{J}$ map, we need the width and $T_\mathrm{J}$ of the stream. $T_\mathrm{J}$ can be readily calculated using the orbits provided by the IAUMDC catalog, but the stream width (i.e., the duration of the shower activity) is not directly provided. Therefore, we look at the original work that published these showers to obtain information on activity duration. For optical streams, the most recent measurements can be obtained from the Cameras for Allsky Meteor Surveillance (CAMS) composite shower look-up table. The most recent release of the table is version 2018-1, which can be accessed at \url{http://cams.seti.org/FDL/data/CAMS-ShowerLookUpTable-v2018-01.txt} \citep{Jenniskens2018}. For radar streams, we use the catalog compiled by \citet{Brown2010}, which is based on the meteor orbits measured by the Canadian Meteor Orbit Radar (CMOR), the largest dataset of its kind.

\begin{figure*}
\includegraphics[width=\textwidth]{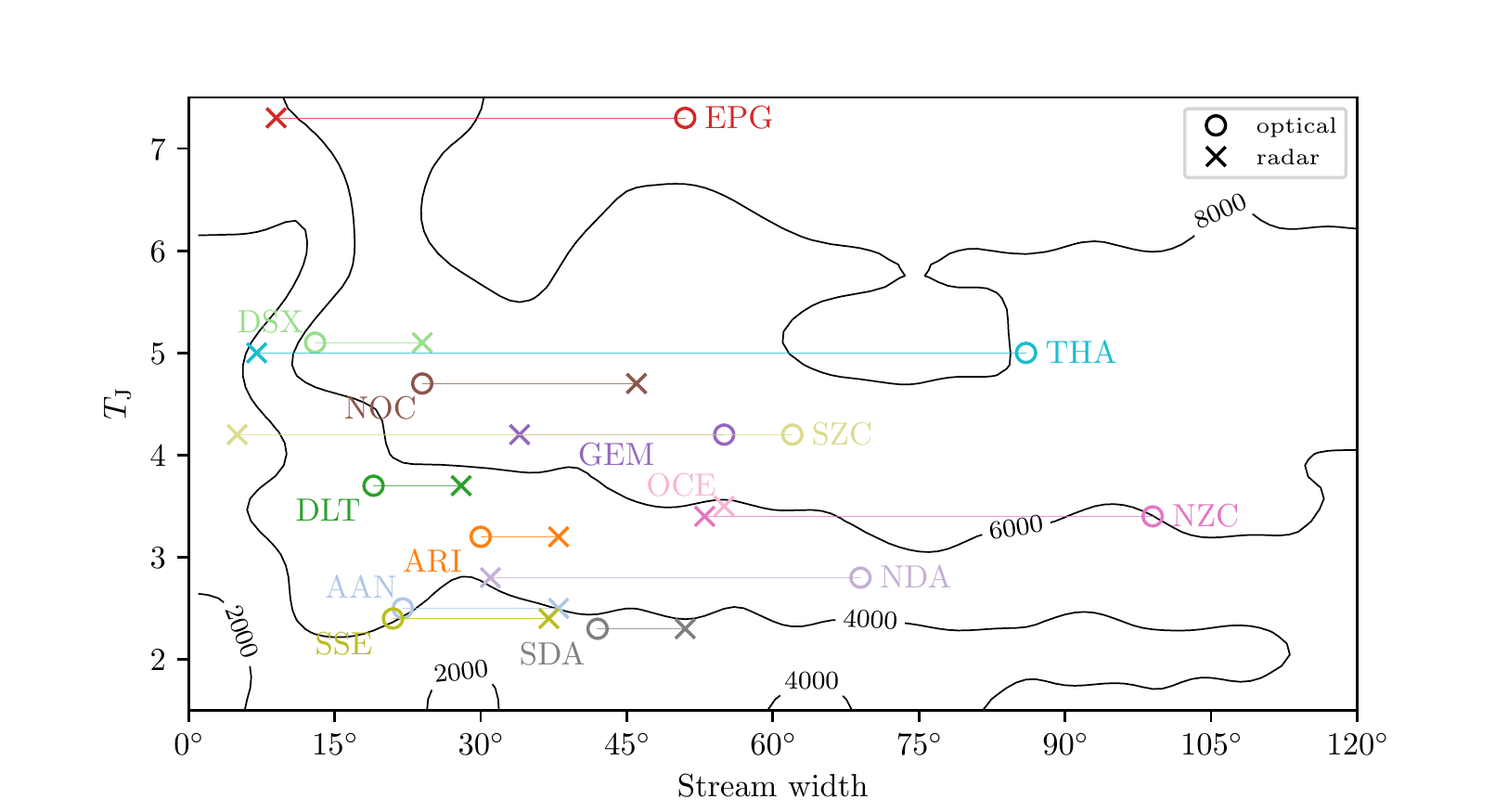}
\caption{Sun-approaching streams with $T_\mathrm{J}>2$ on age-width-$T_\mathrm{J}$ map, assuming ejection at gravitational escape speed. To assist viewing, optical/radar data of the same stream is connected with thin lines. Note that OCE has no optical detection. The Jupyter notebook for this figure is available \href{https://github.com/Yeqzids/near-sun-disruptions/blob/master/nb/age_map.ipynb}{here}. \label{fig:shr-age}}
\end{figure*}

We then overlay the \{width, $T_\mathrm{J}$\} points of the stream-of-interest on the age-width-$T_\mathrm{J}$ map, shown as Figure~\ref{fig:shr-age}. From now on we will stick to the map generated by the gravitational escape ejection model, since it is not essentially different from the one generated by the cometary ejection model, and is more plausible for thermally-driven activity. The map suggests that all these streams have an age of a couple kyr, but as the sanity check by using the 96P/Machholz streams discussed above has shown, these age estimates are somewhat dependent on how the particles are ejected. This will be discussed in greater detail in \S~\ref{sect:known:bias}.

One curious finding is the disagreement on the width of the same stream between optical and radar data: the difference can get up to a factor of 10 in extreme cases (e.g., THA). We note that the GEM optical measurement obtained from the CAMS shower look-up table is also about 7 times wider than earlier measurements shown in \citet{Ryabova2016}. There are two explanations for the difference: uncertainty in the determination of the start and end time of meteor showers may be the culprit \citep[e.g.][\S~4]{Brown2008}, or, the fact that optical and radar systems are sensitive to different size regime of meteoroids. It is difficult to determine which one is more important or if they are equally important. For the latter explanation, there are known examples where the meteoroid size distribution varies across a stream \citep{Campbell-Brown2006a, Ye2014}. Fortunately, the difference does not make a dramatic impact on the age estimates. Even for streams with optical-radar width of a factor of 10, the difference between the resulting age estimates is only up to a factor of 2.

To conclude, even after considering different modes of ejection (instantaneous vs. continuous ejection) and the uncertainty in stream width, the ages of the streams in Table~\ref{tbl:shr} are on the order of a few kyrs with uncertainties likely within a factor of $\sim2$.

\subsection{Possible Biases and Implication} \label{sect:known:bias}

If we compare the derived ages with the dispersion timescale derived in \S~\ref{sect:model:result}, we immediately note that the estimated ages are consistently on the longer side of the predicted dispersion timescale. So where are the young streams?

One plausible explanation is that our consideration of ``established'' showers introduce a bias against young, short-duration, and therefore hard-to-confirm streams. According to Figures~\ref{fig:age-width-tj-cmt} and~\ref{fig:age-width-tj-grav}, streams younger than $\sim1$~kyr are active for less than about a day, and are thus difficult to detect and confirm unless they are strong.

\begin{figure*}
\includegraphics[width=\textwidth]{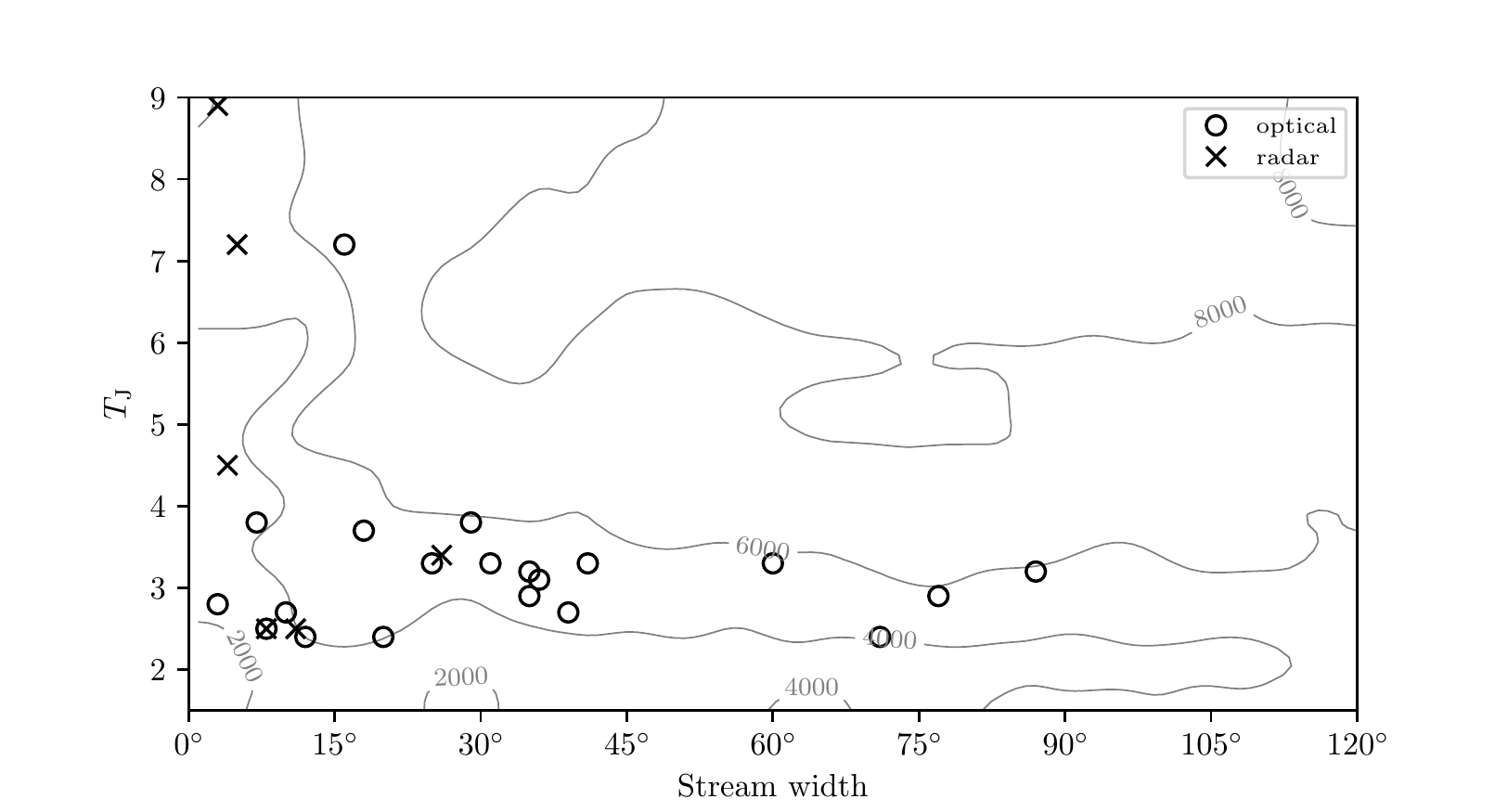}
\caption{Same as Figure~\ref{fig:shr-age}, but showing Working List showers. All of these showers have only either radar or optical detections. The Jupyter notebook for this figure is available \href{https://github.com/Yeqzids/near-sun-disruptions/blob/master/nb/age_map.ipynb}{here}. \label{fig:shr-age-working}}
\end{figure*}

Could the IAUMDC Working List contain some of the young showers? We examine the Working List and plot any qualified showers to the age-width-$T_\mathrm{J}$ map (Figure~\ref{fig:shr-age-working}). There are a few candidates, such as APG ($3^\circ$), DRG ($3^\circ$), TOP ($4^\circ$), and KCT ($5^\circ$), alongside with many low $T_\mathrm{J}$ streams that are only seen by optical systems. It is difficult to say how many short-duration streams we have missed, but the possibility of bias is real. The deployment of optical and radar network across the world \citep[e.g.][]{Janches2015,Pokorny2017, Jenniskens2018, Li2018} will increase the temporal coverage and orbit statistics and enhance our understanding of young streams.

\begin{figure*}
\includegraphics[width=\textwidth]{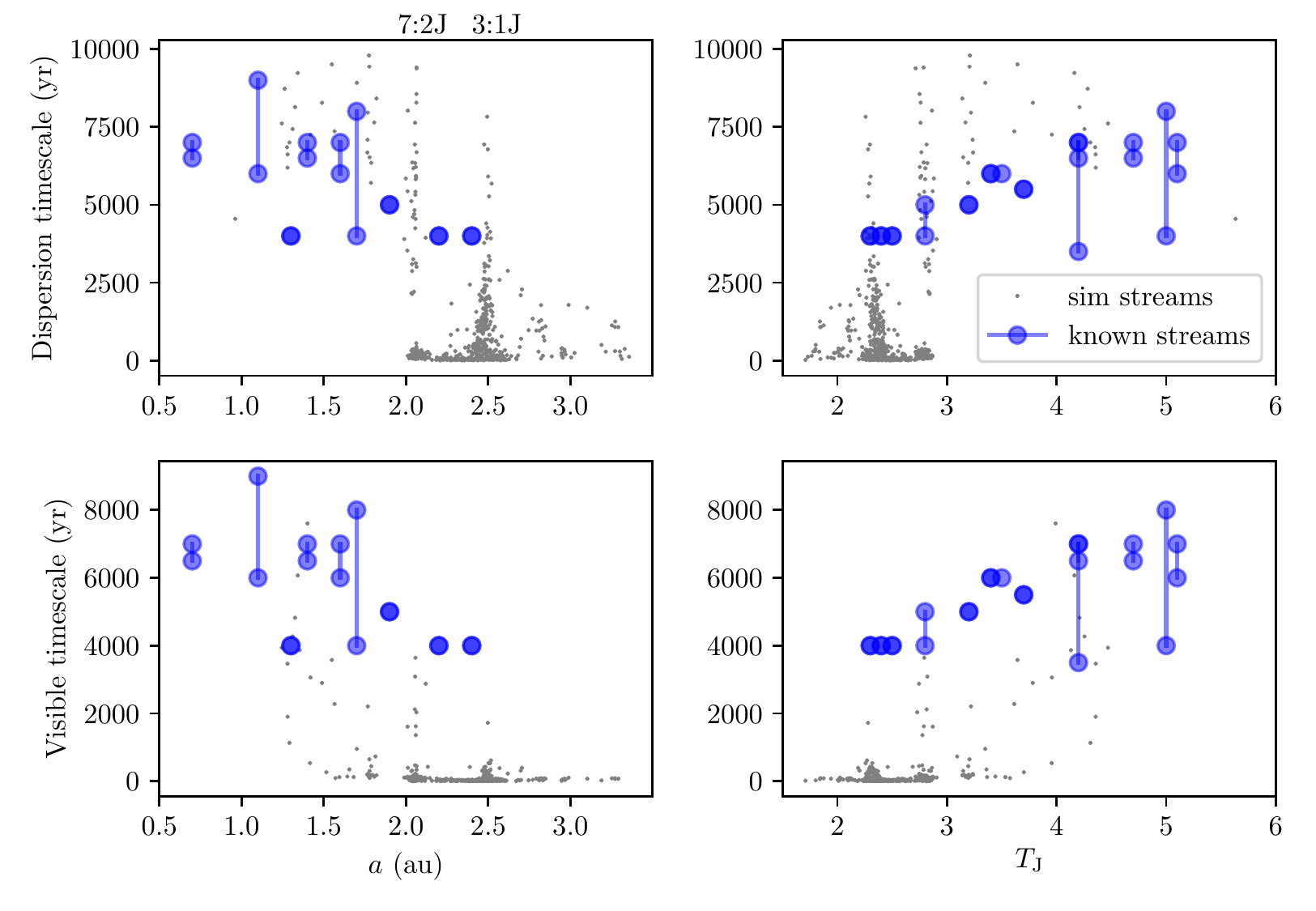}
\caption{Same as Figure~\ref{fig:time-a-tj} but with the estimated age of known streams overlaid. When applicable, extended lines over Y-axis show the difference between radar-based estimates and optical-based estimates. The Jupyter notebook for this figure is available \href{https://github.com/Yeqzids/near-sun-disruptions/blob/master/nb/timescales_established.ipynb}{here}. \label{fig:time-a-tj-known}}
\end{figure*}

Another possibility is that these streams are trapped in resonances and are therefore somewhat immune against dispersion. To test this hypothesis, we plot the semimajor axes of the streams over Figure~\ref{fig:time-a-tj}, with the results shown in Figure~\ref{fig:time-a-tj-known}. We do not find clear concentrations near the 7:2J and 3:1J resonances; in fact, as indicated by the distribution of $T_\mathrm{J}$, about half of these streams are decoupled from Jupiter so that they are immune to the perturbation from Jupiter. On a separate note, this is also consistent with the finding in \S~\ref{sect:model-rate}, that objects from the inner parts of the asteroid belt are more likely to thermally disrupt than those from the outer asteroid belt.

\section{Discussion} \label{sect:disc}

If we compare the model prediction made in \S~\ref{sect:model} to the observations presented in \S~\ref{sect:known}, a clear disagreement is revealed: the model predicts that the likelihood of detecting any thermally-driven stream at the Earth is (visible timescale at the Earth) $\times$ (likelihood of being visible at the Earth) $\times$ (rate of disruption). By plugging in the numbers derived in \S~\ref{sect:model-rate} and~\ref{sect:model:result}, this will be $40$~yr $\times 75\% \times 0.6~\mathrm{kyr^{-1}}=0.02$, or 1 in 50. However, observations have revealed a dozen Sun-approaching streams that are dynamically compatible with asteroids and have no known parents. If all these streams were produced by thermally-driven disruptions, it would require a disruption rate of (number of streams) $\div$ (visible timescale at the Earth) $\div$ (likelihood of being visible at the Earth), or $10\div40\div75\%$~yr$~=300~\mathrm{kyr^{-1}}$, which is $\sim500$ times higher than model prediction. Even if we exclusively focus on the streams with $T_\mathrm{J}>3$, which have a visible probability of $90\%$ and a median visible timescale of $600$~yr\footnote{Details of the calculation can be found in the Jupyter notebook, available \href{https://github.com/Yeqzids/near-sun-disruptions/blob/master/nb/tj_gt_3_streams.ipynb}{here}.}, the likelihood of detection is $600$~yr $\times 90\% \times 0.6~\mathrm{kyr^{-1}}=0.3$. Such a disagreement clearly implies a problem with the assumptions, and the problem may be:

\begin{enumerate}
 \item an underestimation of the rate of thermally-driven disruptions from the NEO population model;
 \item contamination from sources other than thermally-driven disruption; or
 \item a different ejection regime than assumed in our model.
\end{enumerate}

Some educated guesses of the properties of the (ex-)parents may provide clues. Equation~\ref{eq:msmass} shows that the stream mass is proportional to the square of the duration of activity, and it is the deterministic variable for the total stream mass, as (for short-period streams) contributions from other variables are not significant. In the highly idealized example that we discussed in \S~\ref{sect:model:size}, a complete disruption of a 500~m, $10^{11}$~kg asteroid will produce a shower that lasts 5~days. A shower that lasts 10 times longer (more in line with the results in Figure~\ref{fig:shr-age}) would require a 100 times more massive parent (i.e. $10^{13}$~kg in mass), or 2~km in size. For streams that are produced as a result of cometary activity instead of complete disruptions, the parents are larger in size: a close comparison might be the streams that were produced by the 10-km-diameter comet 1P/Halley ($\eta$ Aquarids and Orionids), which is collectively $3\times10^{13}$~kg in mass \citep{Hughes1989}, although we acknowledge that smaller comets could also be responsible for massive streams if they are sufficiently active.

\subsection{Underestimation from the NEO Population Model}

Underestimation from the \citet{Granvik2016} model would imply that the number of thermally-driven disruptions of km-sized bodies over the past $\sim10$~kyr is 1--2 orders of magnitudes more frequent than what \citet{Granvik2016} model predicts. However, \citet{Granvik2016}'s model does otherwise agree well with the observed NEO population, and other models based on a similar modeling approach also show good agreement with the observational data \citep[e.g.][]{Bottke2002, Greenstreet2012}. Therefore, this scenario seems unlikely.

\subsection{Contamination of Streams Produced by Other Mechanisms} \label{disc:comets}

Most known meteoroid streams are produced by comets and therefore one could speculate that Sun-approaching comets may have supplied some of the streams in Table~\ref{tbl:shr}. Compared to thermally-driven disruptions of asteroids, that will convert the entire mass into a meteoroid stream, comets are more ``sustainable'' and only deposit a fraction of its mass to the meteoroid stream as they orbit the Sun (unless they disrupt), and therefore need to be larger in size. However, the streams produced by comets will also be longer-lived since they will be continuously replenished. Especially comets in the Lidov-Kozai resonance like 96P/Machholz will periodically return to low-$q$ zone \citep{Levison2014}, and they can therefore potentially maintain a stream for a very long time.

However, half of the streams in Table~\ref{tbl:shr} are dynamically asteroidal. This scenario also results a large number of comets that is not supported by observations: using the numbers derived in \S~\ref{sect:model:result}, we crudely estimate that a stream has $40/170$--$40/280$ or 1 in 7 to 1 in 4 chance of being detectable at any time, meaning that (statistically speaking) 4--7 comets are needed to produce one detectable stream.

Another potential source of contamination comes from the streams produced at ordinary distances from the Sun that migrate into low-$q$ orbits due to Lidov-Kozai mechanism and Poynting-Robertson drag, as noted by \citet{Wiegert2008}. They found that such process can occur over a timescale of a couple kyr, which is in the range of the ages we derived for the Sun-approaching streams. However, the same study also noted that particles of different sizes have different responses this process: smaller particles are more sensitive to Poynting-Robert drag while larger particles are largely immune, which implies that these ``immigrants'' should largely be only detectable in radio wavelengths, a prediction that is incompatible with observations.

\subsection{Prolonged Disruption Phase} \label{disc:prolonged}

For the case of the Machholz complex which we discussed in \S~\ref{sect:age:map}, we see that estimates by instantaneous ejection model can underestimate the age of continuously replenished streams by a factor of several. The latter streams are also longer-lived since their cores will be replenished and will stay compact over time. Asteroids that have completely disintegrated will no longer be able to replenish their streams; however, if their disruption phase span over multiple orbits, they will be able to maintain their streams longer than what our instantaneous model predicts.

In order to maintain 10 Sun-approaching streams, the visible timescale of each stream would need to be $10/0.6/75\%~\mathrm{kyr^{-1}}=22$~kyr, which is 500 times longer than the prediction made for instantaneous ejection. Assuming the relation between visible timescale and dispersion timescale is fixed, each parent would need to ``shepherd'' its stream over a 10--20~kyr timescale. The timescale shifts to the shorter end if there are fewer streams to maintain: for 1 stream the timescale becomes as short as 1--2~kyr.

Besides meteor observations, another observational implication of the ``prolonged-disruption'' theory is this: instead of a gigantic, one-time near-Sun explosion event that happens once every $\sim2$~kyr, observers are more likely to see a handful Sun-approaching asteroids that will only release a fraction of its mass, and therefore only mildly brighten during their perihelion passages. Existing observational evidence is consistent with this theory: in the past several decades, a couple of periodic Sun-approaching objects have been discovered by the Solar and Heliospheric Observatory (SOHO). These objects are all designated as comets since most objects detected at such distances are sun-grazing comets \citep{Knight2010}, even though most, if not all these periodic objects never display any coma or tail. Here we note that objects are only detectable by SOHO when they get to $V=8$ or brighter, since the limiting magnitude of SOHO is about $V=8$. Therefore, these SOHO ``comets'' must either be very large, or are actively producing dust. However, these SOHO ``comets'' have never been detected by ground-based NEO surveys which are sensitive down to $V=21$ \citep{Jedicke2015}, which suggests that they are no larger than typical NEOs detected by surveys ($\sim0.1$ to a few km in sizes). This also suggests that these objects are apparently inactive when away from perihelion even though they are inside the sublimation distance of water ice, the dominant volatile species in comets, seemingly implying that these are asteroids. In fact, dedicated observations of one such objects, 322P/SOHO, have shown characteristics consistent with asteroids \citep{Knight2016}.

\begin{figure}
\includegraphics[width=0.5\textwidth]{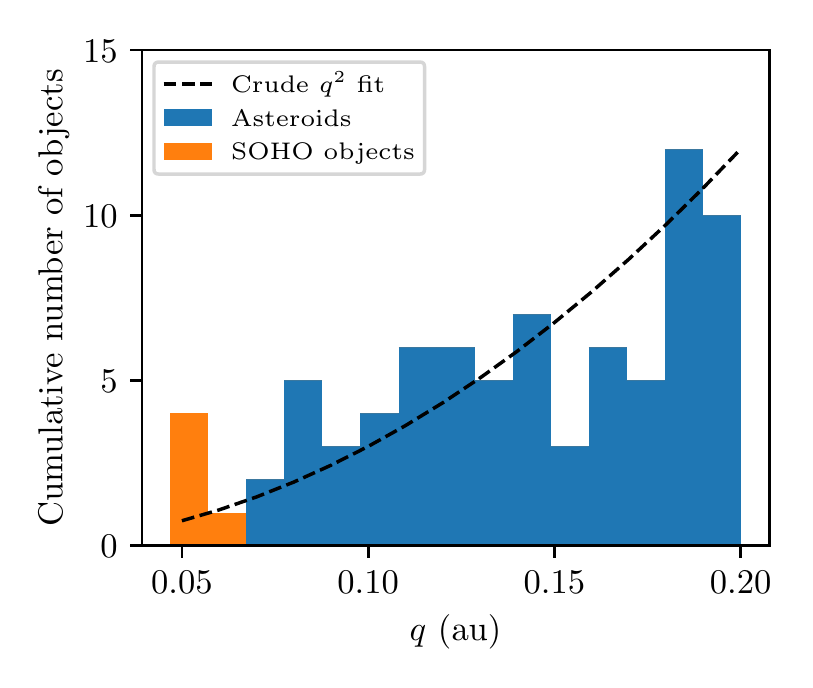}
\caption{Distribution of perihelion distance $q$ of objects with $q<0.2$~au and $T_\mathrm{J}>2$. Objects found by SOHO (meaning that they will get to $V<8$ during perihelion) and ground-based NEO surveys are shown in different colors. Also shown is a $q^2$ curve that gauges the number density of objects as a function of $q$. The Jupyter notebook for this figure is available \href{https://github.com/Yeqzids/near-sun-disruptions/blob/master/nb/soho_objects.ipynb}{here}. \label{fig:soho}}
\end{figure}

Figure~\ref{fig:soho} shows the orbital distribution of all known objects with $q<0.2$~au and $T_\mathrm{J}>2$, with objects found by SOHO and ground-based NEO surveys in different colors. Here we clearly see a transition at 0.06~au, the distance that thermally-driven disruption is predicted to occur by \citet{Granvik2016}. We also overlay a $q^2$ curve to crudely match the histogram, which shows the number density stays constant throughout the region and argues against contamination from other small body populations (e.g. comets). Therefore, we conclude that at least some of the periodic SOHO ``comets'' might, in fact, be asteroids that are being destroyed.

The $q^2$ fit also provides some clues into the timescale of the disruption. If the timescale is short, then we should see no more than a few objects in this regime, since they are removed quite efficiently once they reach the disruption distance; if the timescale is long, then we should see a ``piled-up'' of objects in this regime. The fact that the distribution crudely agrees with the fit suggests that the timescale is unlikely to be too much different from the derived disruption rate of 0.6~$\mathrm{kyr^{-1}}$, which is also in line with the timescales we derived at the beginning of this section, although we caution that the small statistics effectively limits our discussion to the order-of-magnitude level.

\section{Conclusions}

It is quite possible that Sun-approaching meteoroid streams are fed by multiple sources. Some of $2<T_\mathrm{J}<3$ streams could have been produced by short-period comets instead of asteroids, while the possibility that thermally-driven disruption of asteroids could be a relatively lengthy process may also contribute to the number of Sun-approaching streams being seen.

The hypothesis that thermally-driven disruption is a lengthy process also implies that such a process might be observable on the current Sun-approaching asteroids, with SOHO objects as prominent examples. \citet{Granvik2016} predicts that thermally-driven disruption can occur on small, dark asteroids at slightly larger distances, possibly up to $\sim0.4$~au. The brightening effect on asteroids at a few tenths of an au will be less pronounced and would be difficult for SOHO and other coronagraphs to detect. Ground-based observers, on the other hand, could have sufficient sensitivity to detect any brightening but have difficulties with small solar elongations. However, space probes operating close to or inside the orbit of Mercury such as MESSENGER, BepiColombo or Parker Solar Probe will have a chance to test this hypothesis.

Our work offers some lessons for understanding solid-body disruptions in other planetary systems. Although we are yet to be able to directly image small bodies in exoplanetary systems, signatures of disrupting small bodies have been found in a variety of planetary systems \citep[e.g.][]{Montgomery2012, Kiefer2014, Vanderburg2015, Rappaport2016, Xu2016, Xu2018}. The direct exoplanetary equivalents of the debris streams investigated in this work (likely at the level of $10^{-10}$--$10^{-12}$ Earth mass) is beyond the detection capability of current techniques, but disruptions of larger bodies that are potentially detectable by current techniques should be governed by the same physics. Our finding of prolonged disintegration of Sun/star-approaching rocky bodies suggests that the observing window for such events is long, and may partly explain the common occurrence of such phenomenon in exoplanetary systems. 




\acknowledgments

The authors thank Peter Brown for his contribution to an early version of the draft and Paul Wiegert for access to computational resource. Q.-Z. Ye is supported by the GROWTH project funded by the National Science Foundation under Grant No. 1545949. M. Granvik is supported by Grant No. 299543 from the Academy of Finland. This work was made possible by the facilities of the Shared Hierarchical Academic Research Computing Network (SHARCNET:www.sharcnet.ca) and Compute/Calcul Canada. We extend our thanks to the American Astronomical Society's Division for Planetary Science for holding its annual meeting, which provides an opportunity for the authors to meet in person and to fight their procrastination.

\facilities{SHARCNET}
\software{Astropy \citep{Robitaille2013}, Jupyter Notebooks \citep{Kluyver2016}, Matplotlib \citep{Hunter2007}, MERCURY6 \citep{Chambers1999}, NumPy \citep{Walt2011}}

\end{CJK*}
\bibliographystyle{aasjournal}
\bibliography{ms}



\end{document}